\documentclass[twocolumn]{aastex631}

\usepackage{gensymb}
\usepackage{mhchem}
\usepackage{float}

\begin{document}


\title{Frequency Comb Calibrated Laser Heterodyne Radiometry\\ for Precision Radial Velocity Measurements}

\author{Ryan K. Cole}
\altaffiliation[]{Both authors contributed equally to this work.}
\affiliation{Time and Frequency Division, National Institute of Standards and Technology\\ Boulder, Colorado, USA}
\affiliation{Department of Physics and Astronomy, Bates College, Lewiston, Maine, USA}%

\author{Connor Fredrick}%
\altaffiliation[]{Both authors contributed equally to this work.}
\affiliation{Time and Frequency Division, National Institute of Standards and Technology\\ Boulder, Colorado, USA}
\affiliation{Department of Electrical, Computer, and Energy Engineering, University of Colorado Boulder\\ Boulder, Colorado, USA}

\author{Winter Parts}%
\affiliation{Department of Physics and Astronomy, The Pennsylvania State University\\ University Park, Pennsylvania, USA}
\affiliation{Center for Exoplanets and Habitable Worlds, The Pennsylvania State University\\ University Park, Pennsylvania, USA}

\author{Max Kingston}
\affiliation{Department of Physics and Astronomy, Carleton College, Northfield, Minnesota, USA}

\author{Carolyn Chinatti}
\affiliation{Department of Physics and Astronomy, Carleton College, Northfield, Minnesota, USA}

\author{Josiah Tusler}
\affiliation{Department of Physics and Astronomy, Carleton College, Northfield, Minnesota, USA}

\author{Suvrath Mahadevan}%
\affiliation{Department of Physics and Astronomy, The Pennsylvania State University\\ University Park, Pennsylvania, USA}
\affiliation{Center for Exoplanets and Habitable Worlds, The Pennsylvania State University\\ University Park, Pennsylvania, USA}

\author{Ryan Terrien}%
\affiliation{Department of Physics and Astronomy, Carleton College, Northfield, Minnesota, USA}

\author{Scott A. Diddams}%
\affiliation{Department of Electrical, Computer, and Energy Engineering, University of Colorado Boulder\\ Boulder, Colorado, USA}
\affiliation{Department of Physics, University of Colorado Boulder, Boulder, Colorado, USA}
\affiliation{Time and Frequency Division, National Institute of Standards and Technology\\ Boulder, Colorado, USA}

\email{rcole@bates.edu, scott.diddams@colorado.edu}

\begin{abstract}

Disk-integrated observations of the Sun provide a unique vantage point to explore stellar activity and its effect on measured radial velocities. Here, we report a new approach for disk-integrated solar spectroscopy and evaluate its capabilities for solar radial velocity measurements. Our approach is based on a near-infrared laser heterodyne radiometer (LHR) combined with an optical frequency comb calibration, and we show that this combination enables precision, disk-integrated solar spectroscopy with high spectral resolution ($\sim$800,000), high signal-to-noise ratio ($\sim$2,600), and absolute frequency accuracy. We use the comb-calibrated LHR to record spectra of the solar \ion{Fe}{1} 1565 nm transition over a six-week period. We show that our measurements reach sub-meter-per-second radial velocity precision over a single day, and we use daily measurements of the absolute line center to assess the long-term stability of the comb-calibrated LHR approach. We use this long-duration dataset to quantify the principal uncertainty sources that impact the measured radial velocities, and we discuss future modifications that can further improve this approach in studies of stellar variability and its impact on radial velocity measurements.

\end{abstract}


\section{\label{sec:intro}Introduction}

Radial velocity (RV) measurements are a central technique used to detect and characterize exoplanets. RV measurements enable the discovery of non-transiting exoplanets and the confirmation of transiting exoplanet candidates. RV measurements also provide dynamical constraints on the exoplanetary mass, which is vital for understanding an exoplanet's internal and atmospheric structure \citep{2019ApJ...885L..25B}. Currently, state-of-the-art spectrographs are able to measure RV shifts with an instrumental precision better than 1~m/s \citep{2021A&A...645A..96P,2016SPIE.9908E..6TJ,2016SPIE.9908E..6FT,2016SPIE.9908E..7HS,2016SPIE.9908E..70G, ford2024earthsreachevaluationstrategies} equivalent to measuring fractional shifts in the optical frequency at the level of a few parts per billion. At this level of precision, the primary challenge for exoplanet detection and characterization is the variability of the stellar spectrum itself \citep{2016PASP..128f6001F,2021arXiv210714291C}, driven by changes in the stellar atmospheric flows and the dynamic stellar magnetic field \citep{2021arXiv210406072M}. This activity-related variability manifests, to varying degrees, as an apparent RV shift, and can contaminate (or mask altogether) any exoplanet-induced RV signal. 

Observations of the Sun provide a vital benchmark for techniques aiming to decouple activity and exoplanet-related RV signals. In particular, ``Sun-as-a-star" observations integrate light from across the solar disk to mimic unresolved observations of stars. The measured RV from Sun-as-a-star observations can be compared to the known configuration of the solar atmosphere at a given time (as monitored, for example, by the solar Dynamics Observatory), which provides the ``ground truth" about the activity state. As such, numerous RV spectrographs have solar feeds that enable this type of Sun-as-a-star monitoring, including NEID, HARPS-N, EXPRES, and KPF \citep{2022AJ....163..184L,2016SPIE.9912E..6ZP,2023PASP..135l5002R, llama2024lowellobservatorysolartelescope}. \cite{zhao2023extreme} provides a recent summary of many of these instruments and their current capabilities for tracking subtle activity signals. Additionally, several dedicated helioseismology observatories also record precision Sun-as-a-star RVs in order to monitor and study solar oscillations, including BiSON \citep{chaplin1996bison, hale2016performance} and GONG \citep{harvey1996global}.

Here, we describe a new instrument for disk-integrated solar spectroscopy based on laser heterodyne radiometry (LHR) \citep{parvitte_infrared_2004, menzies2005laser}. In LHR, thermal light is combined with light from a wavelength tunable laser and interfered on a photodetector. The resulting heterodyne signal generated between the thermal and laser light is proportional to the power of the thermal light within a narrow frequency range around the tunable laser. Tuning the laser frequency thus gives a measure of the thermal light spectrum within the scan range of the laser. LHR is a well-known approach for atmospheric remote sensing \citep[e.g.][]{tsai_atmospheric_2012,deng_development_2021, sappey_development_2021, stupar_fully_2008, rodin_high_2014, weidmann_high-resolution_2009, wilson_miniaturized_2014, hoffmann_thermal_2016, bomse_precision_2020}, and several studies have also used the approach to measure solar absorption transitions or spectra of other astronomical sources \citep[e.g.][]{goldstein_absolute_1991, sappey_passive_2020, fredrick_thermal-light_2022, nieuwenhuijzen_optical_1970, kostiuk_remote_1983, peyton1975infrared, sonnabend2002tuneable}. However, to the best of our knowledge, LHR has never been implemented or tested as an approach for long term, continuous, Sun-as-a-star observations such as those required to explore the links between solar variability and precision radial velocity measurements. 

LHR contributes a fundamentally different measurement principle when compared to the other solar observing instruments described above. As a laser-based approach, LHR can be directly combined with well known tools for precision laser spectroscopy such as frequency combs \citep{fredrick_thermal-light_2022} or modulation techniques \citep{martin2018wavelength}, and can also reach very high spectral resolution ($>10^6$) without moving components or diffractive optics. These benefits could prove LHR to be a powerful tool for solar spectroscopy to complement existing high-performance solar spectrographs in studies of solar variability. 

In this paper, we describe the design and operation of a frequency comb calibrated LHR system used for long-term measurements of the solar \ion{Fe}{1} 1565~nm transition. Through the unique combination of LHR and a frequency comb calibration, our measurements reach sub-m/s radial velocity precision within a single day, and we use continuous measurements of the absolute line center over a period spanning $\sim$6 weeks to explore the long-term stability of our approach. These results, along with a thorough description of the relevant uncertainty sources, help inform the precision and accuracy limits of LHR-based solar spectroscopy along with future modifications that could improve the utility of the approach for studies of solar variability and radial velocity measurements.

\section{\label{sec:inst}Instrument Description}
\begin{figure*}[t]
\includegraphics[width=\linewidth]{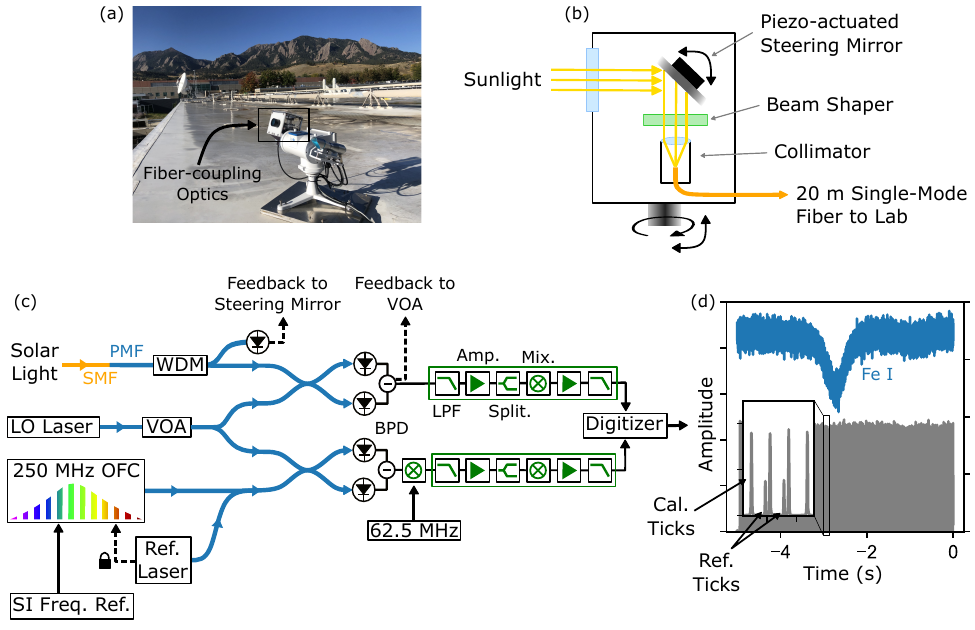}
\caption{Schematic of the frequency comb-calibrated LHR system currently in operation at NIST Boulder. Panel (a) shows the rooftop-mounted solar tracker and panel (b) highlights the fiber-coupling optics mounted to the solar tracker. Panel (c) shows the fiber optic configuration and RF power detection circuit (green box) used to measure the 1565 nm solar Fe I transition and frequency calibration ``ticks" as the LO laser is scanned across the solar transition (panel (d)). Light from an additional reference laser locked to the frequency comb generates an additional pair of ``reference ticks" that are used to determine the absolute frequency of the measured spectrum. The frequency comb is referenced to a NIST-calibrated hydrogen maser that is directly traceable to the SI second. All components in Panel (c) are located in a rooftop laboratory adjacent to the solar tracker. OFC: optical frequency comb, WDM: wavelength division multiplexer, VOA: variable optical attenuator, SMF: single-mode optical fiber, PMF: polarization-maintaining optical fiber, BPD: balanced photodetector, LPF: low-pass filter, Amp: amplifier, Split: power splitter, Mix: RF mixer. 
\label{fig:expt}}
\end{figure*}

Our frequency comb calibrated LHR approach has been described in prior works from our group \citep{fredrick_thermal-light_2022, cole_precision_2023}. Here, we describe recent modifications to the instrument design that are intended to improve the precision and accuracy of the instrument for solar spectroscopy. Figure \ref{fig:expt} shows a schematic of the comb-calibrated LHR system and solar tracking telescope currently in operation at NIST Boulder, and the sections below describe the major components. 

\subsection{\label{sec:tracking}Solar Tracking and Fiber Coupling}
The optical and mechanical design of the solar tracking system is critical to mitigate unintended biases in the measured spectra and to enable Sun-as-a-star observations in which the solar disk is intentionally unresolved. Imperfections in solar tracking (e.g. pointing drift away from the center of the solar disk) can lead to bias in measured spectra due to the introduction of rotational Doppler shifts arising from the preferential coupling of light from one region of the solar disk. Our approach addresses this challenge in two ways: by employing a dual-stage solar tracking approach, and by optimizing the antenna pattern of our solar telescope to match the solar disk. 

We couple solar light into single-mode fiber in a weatherized enclosure mounted to a commercial solar tracker (EKO STR-22G), identical to the model employed on the NEID solar telescope \citep{lin2022observing}. In addition, we use a piezo-actuated steering mirror (TEM Messtechnik Fiberlock) housed in the weatherized enclosure  to correct for minor imperfections in the commercial solar tracker. The feedback signal for the steering mirror is derived from the fiber-coupled solar power at wavelengths below $\sim$1000~nm. These wavelengths are split from the desired 1565~nm signal in a low-loss fiber wavelength division multiplexer (WDM) and measured on a silicon photodetector located in a rooftop laboratory nearby the solar tracker (see Figure \ref{fig:expt}). Pointing deviations away from the center of the solar disk result in a decrease in the fiber-coupled solar intensity, and these deviations are then corrected by the steering mirror that maximizes the fiber-coupled intensity. 

In addition to the two-stage solar tracking, we employ a custom refractive beam shaping optic between the steering mirror and the fiber coupler to transform the Gaussian mode of the single-mode fiber to a flat-top profile in the far field. This flat top profile (the ``antenna pattern") provides a uniform integration of light from across the solar disk to emulate an observation in which the Sun is unresolved. Additionally, the flat top profile balances signal-to-noise ratio and pointing sensitivity when compared to a Gaussian antenna pattern that preferentially couples light from the center of the solar disk \citep{fredrick_thermal-light_2022}. Section \ref{sec:pointingunc} provides a further description of this flat-top profile and its effect on our measurements. 

\subsection{\label{sec:lhr}LHR and Frequency Comb Calibration}

Fiber-coupled solar light travels through 20~m of single mode fiber (SMF~28) to an adjacent rooftop laboratory housing additional LHR instrumentation. Here, the solar light is  passed through a fiberized polarizing beam splitter and combined with LO light from a 1565~nm distributed feedback laser (linewidth $\sim$ 3~MHz) that is temperature tuned over the target absorption transition. The solar and LO light are combined in a polarization-maintaining fiber optic coupler and mixed on a balanced InGaAs photodetector (Thorlabs PBD465C). The total bandwidth of the recieved solar light is limited by the InGaAs responsivity to $\sim$1000-1700~nm. The radio frequency (RF) photodetector output is passed to a RF power detection circuit (described below), and the DC monitor output is used to feed back to a variable optical attenuator (VOA) that stabilizes the LO laser power throughout each scan. Heterodyne detection is a single-mode process, thus the RF heterodyne signal is generated only from the component of the solar light matching the polarization state of the LO. However, in this case, the exact polarization state of the solar light that we measure is not well defined owing to the long length of non-polarization-maintaining single-mode fiber connecting the solar telescope to the LHR instrumentation. 

The RF output of the photodetector is passed through a low pass filter (LPF) that sets the spectral resolution of the spectrometer as twice the filter cutoff frequency. The filtered heterodyne signal is amplified and rectified using the combination of a 0$\degree$ power splitter and double-balanced mixer. The mixer output is a DC signal that is directly proportional to the heterodyne signal power, and thus proportional to the solar optical power within the frequency range defined by the initial low pass filter and centered on the LO laser frequency. The mixer output signal is amplified and passed to a final low pass filter before being digitized on an oscilloscope. 

In a second branch, light from the scanning LO laser is simultaneously interfered with light from a stabilized, Er:fiber mode-locked frequency comb ($f_r=250$~MHz) on a balanced photodetector. The resulting heterodyne signal is mixed with a synthesized 62.5~MHz tone and filtered using a 2~MHz low pass filter. The remainder of the RF power detection circuit is the same as described above. The output of this process is a series of calibration ``ticks" that register each time the scanning LO laser is 62.5~MHz from a frequency comb mode. As such, the addition of the 62.5~MHz tone effectively doubles the density of frequency calibration points, and the resulting calibration grid is spaced by exactly $f_r/2=125$~MHz \citep{fredrick_thermal-light_2022, jennings_frequency_2017}. 

In the frequency domain, accounting for the increased density of calibration points, the calibration grid follows the modified comb equation
\begin{equation}\label{eq:comb}
f_n = \left (n+\frac{1}{2}\right )\left (\frac{f_r}{2}\right ) + f_o,
\end{equation}
where $f_n$ is the frequency of the $n^{\rm th}$ calibration tick, $f_r$ is the repetition rate, and $f_o$ is the carrier-envelope offset frequency of the comb. Our comb is referenced to a NIST-calibrated hydrogen maser, and thus the resulting calibration grid is directly traceable to the SI second with fractional uncertainty of a few parts in $10^{13}$ or better. Further details regarding uncertainty in this calibration process are discussed in Section \ref{sec:calunc}.

One subtlety of the frequency calibration is that the calibration grid only provides knowledge of the relative frequency of the measured transition. Specifying the absolute optical frequency requires knowledge of the integer index $n$ for a specific calibration tick. To address this ambiguity, we phase lock a second continuous-wave (CW) laser to the frequency comb at a known offset frequency, and we combine light from this ``reference" CW laser with the comb light at the input of the calibration channel. The addition of this CW laser light in the calibration process registers as an additional pair of reference tick marks, again spaced by $f_r/2=125$~MHz, but offset from the regular calibration grid by the phase lock offset frequency (see Fig. \ref{fig:expt}). These reference tick marks identify a single calibration tick in each LO laser scan. We can unambiguously determine the index of this calibration tick by measuring the frequency of the stabilized CW laser with a commercial wavemeter (here a Bristol~621, accuracy $\pm$0.2 ppm at 1565~nm) and solving Eq. \ref{eq:comb} with the known CW laser and lock frequencies. This process fixes the index $n$ for a single tick mark in our calibration grid, and thus enables absolute, comb-referenced optical frequency measurements with our LHR system.

\subsection{\label{sec:datareduc}Post Processing and Frequency Calibration}

The comb-calibrated LHR approach simultaneously records a DC signal proportional to the solar spectrum as well as a series of comb-calibration ticks for each measured spectrum. The solar signal is measured relative to the dark signal (i.e. the signal level with no incident solar power). We zero-point correct each measured spectrum by subtracting the average dark signal level measured daily before and after data collection. The dark signal level varies each day at the level of a few parts in $10^4$, however these variations only result in a small DC offset that does not affect line center measurements. We also rescale each measured spectrum using a transfer function that relates the measured heterodyne signal power to an optical power \citep{fredrick_thermal-light_2022}. This transfer function is determined by recording the heterodyne signal generated between the LO laser and an amplified spontaneous emission (ASE) source across a range of ASE powers, and is nearly linear for the optical powers involved in solar measurements. The rescaling process relates the measured DC signal to a corresponding optical power and also compensates for any nonlinearity in the RF power detection process. 

Before calibrating the temporal axis of the measured signal to the comb-referenced optical frequency grid, we first account for the differential group delay between the solar and comb-calibration branches that could manifest as a frequency shift after the calibration process. To account for this group delay, we measure the phase response through the final amplification and low pass filtering stages in both the solar and comb-calibration branches using a vector signal analyzer (HP~89410A). We negate the group delay and bring both signals into the same temporal frame by multiplying each signal by the inverse of the corresponding phase response in the frequency domain. Further details about this process and its associated uncertainty are discussed in Section \ref{sec:calunc}.

To calibrate the frequency of each measured spectrum, we fit each comb-calibration tick to determine its centroid. We use these calibration points along with the known frequency spacing between each tick mark ($f_r/2=125$~MHz) to construct a time-to-frequency calibration function. We interpolate this calibration function using a $\mathrm{2^{nd}}$ order polynomial to transform the temporal axis of each measurement to the comb-referenced frequency grid in the laboratory frame. The final step in the frequency calibration process is to apply a barycentric correction \citep{wright_barycentric_2020, kanodia_python_2018} that accounts for the relative motion between our observatory and the Sun by transforming the laboratory frequency grid to a grid that is at rest with respect to the solar system's barycenter. Uncertainties related to the frequency calibration process are discussed below.

\section{\label{sec:results}Results}

Using the approach described above, we used the comb-calibrated LHR to measure a solar \ion{Fe}{1} transition near a vacuum wavelength of 1565.28~nm \citep{ryabchikova2015major, peterson2014new} over a period of more than six weeks from 2023 September 21 to 2023 November 5. Spectra were recorded each day beginning at sunrise using an automated data collection routine. Each individual spectrum was recorded over a LO laser scan lasting approximately 5~s spanning a frequency range of $\sim$34~GHz. The measured heterodyne signal was filtered using a $\sim$115~MHz low pass filter, which results in a spectral resolution of 230~MHz or a resolving power ($\lambda/\Delta\lambda$) of approximately 800,000. For a thermal source with the temperature of the Sun, the optical power contained in this resolution bandwidth is $\sim$7.5~pW. The final low pass filter in the RF power detection chain results in an effective averaging time of $\sim$0.28~ms, which yields $\sim$100 independent samples per resolution bin. 

The signal-to-noise ratio (SNR) for each measured spectrum is $\sim$39 at the continuum level. The SNR is limited by the shot noise of our LO laser, and in this limit the LHR SNR obeys a simple expression \citep{zmuidzinas_thermal_2003} 
\begin{equation}\label{eq:snr}
SNR = \frac{\eta \langle n \rangle}{1 + \eta \langle n \rangle}\sqrt{\Delta \nu \tau},
\end{equation}
where $\langle n \rangle$ is the mean photon occupancy given by the Planck distribution (a reasonable approximation for the solar spectrum near 1565~nm; \cite{iqbal2012introduction}), $\Delta \nu$ is the optical bandwidth (here 230~MHz), $\tau$ is the averaging time ($\sim$0.28~ms), and $\eta$ is an efficiency factor that accounts for all loss mechanisms between the Sun and our photodetector. Assuming this equation (with $T_\odot = 5772$~K \citep{prvsa2016nominal}), our observed SNR implies an efficiency $\eta$ of approximately 0.7, which includes factors such as loss in our fiber optic components, detector quantum efficiency, atmospheric losses, mismatch between our antenna pattern and the solar disk, etc. \newline

\subsection{\label{sec:resultlineshape}Line Shape Measurements}

\begin{figure}[t]
\includegraphics[width=\linewidth]{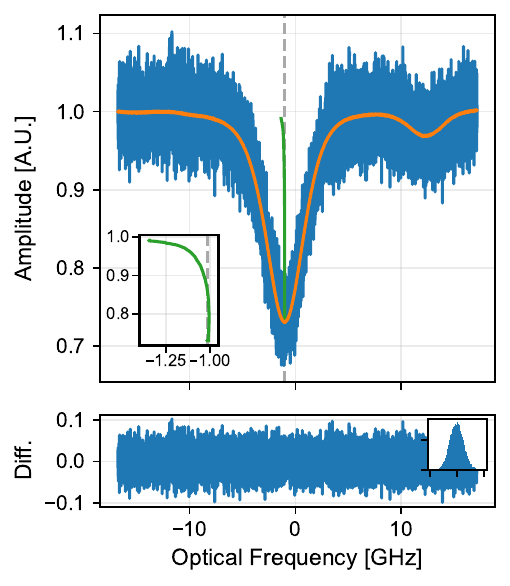}
\caption{Measured solar spectrum from 2023 October 7. The optical frequency is specified relative to 191.52759375(4) THz. The blue curve shows a single 5~s measurement while the orange curve show the spectrum after averaging over the full day. The green curve shows the line bisector determined from the averaged spectrum. Telluric lines have been subtracted using the procedure described in Sec. \ref{sec:telluricunc}, and the vertical dashed line indicates the expected position of the \ion{Fe}{1} line specified by the VALD3 database \citep{ryabchikova2015major, peterson2014new}.
\label{fig:lineshape}}
\end{figure}

Figure \ref{fig:lineshape} shows a representative measurement of the solar \ion{Fe}{1} transition from 2023 October 7. The figure shows both a single 5~s measurement as well as the final spectrum after averaging over the full measurement period. The SNR grows with averaging, and exceeds 2,600 after averaging for $\sim$7 hours. The \ion{Fe}{1} transition is flanked by two telluric lines (not shown in Figure \ref{fig:lineshape}) and one additional, unidentified solar transition. The telluric lines are both due to atmospheric water vapor absorption, and we fit and subtract these lines from our measured spectra using a HITRAN2020-based \citep{gordon_hitran2020_2021} absorption model for a multilayer atmosphere. Further details on our telluric correction protocol are discussed in Section \ref{sec:telluricunc}. 

Although the majority of this paper is focused on the Doppler shifts and corresponding radial velocities determined from our measured spectra, the high SNR and high spectral resolution of our line shape measurements can provide a unique vantage point to explore solar activity through its effect on the line shape. As an example of these line shape effects, Figure \ref{fig:lineshape} shows the bisector of the measured absorption transition which indicates a pronounced asymmetry in the line shape. The shape of this bisector curve can provide information about convection and granulation patterns in the solar atmosphere \citep{dravins_d_solar_1981}, while the absolute line center is influenced by both a convective blue shift \citep{lohner2018convective} and a gravitational red shift \citep{gonzalez_hernandez_solar_2020}. Although we do not analyze these line shape effects in detail in this paper, future studies will leverage the high SNR and high resolution of our line shape measurements to explore indicators of stellar activity in the infrared.

\subsection{\label{sec:resultrvs}Radial Velocity Measurements}

\begin{figure*}[p!]
\includegraphics[width=\linewidth]{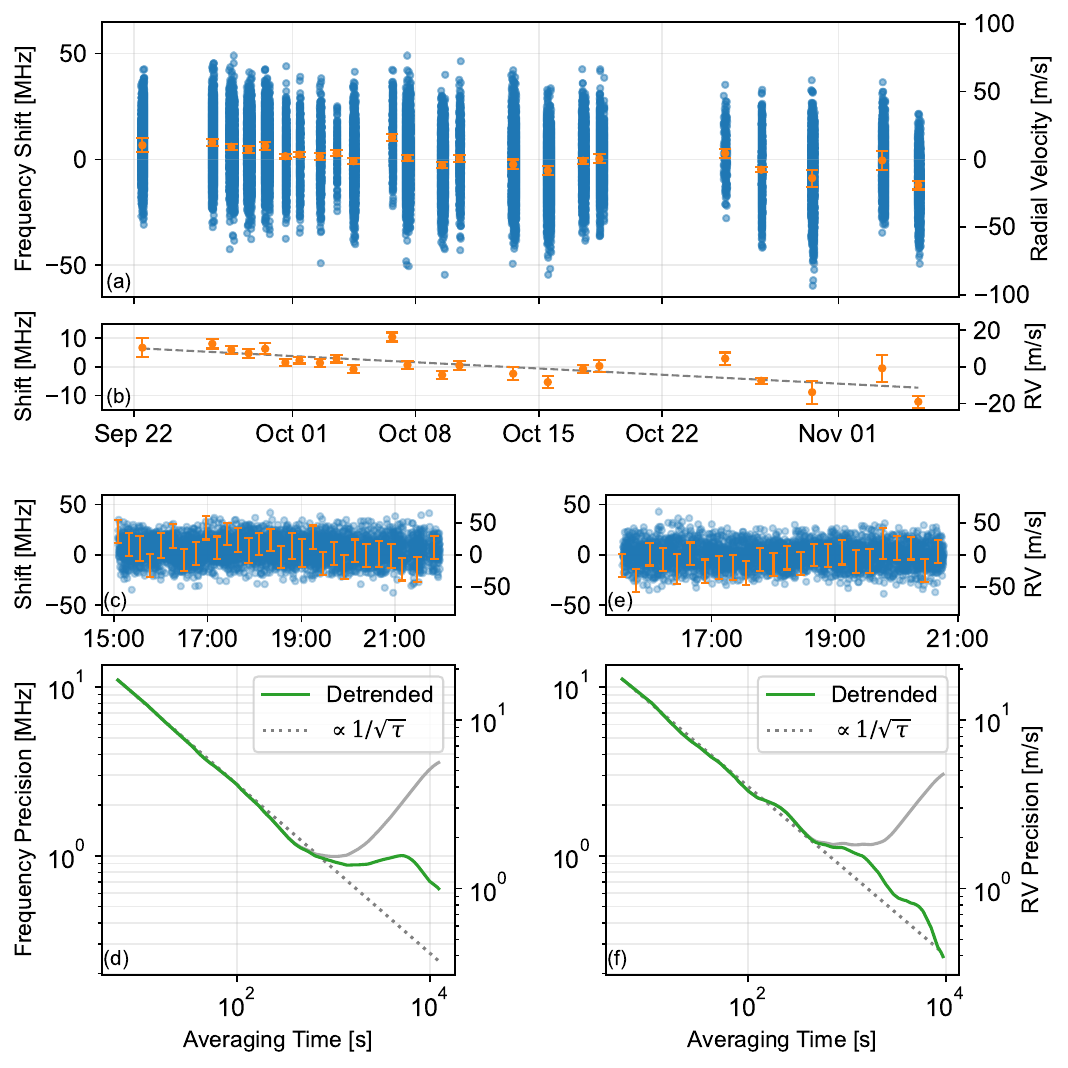}
\caption{Radial velocities measured using the comb-calibrated LHR system from 2023 September 21 to 2023 November 5. Panel (a) shows the individual RV measurements (blue points) and the daily average RV (orange points) for each day. Panel (b) highlights the daily average RV measurements, which indicate a slow drift by $\sim$310~kHz/day. Panels (c) and (d) show the RV measurements for a representative day (2023 September 28) and corresponding Allan deviation showing the frequency precision of our measurements before (gray) and after (green) detrending to correct for the pointing-induced diurnal drift.  Error bars are shown for every 150 measurements for clarity. Panels (e) and (f) show RV measurements for 2023 October 17. All times are specified in UTC. All error bars are calculated using the methods described in Section \ref{sec:uncert}.
\label{fig:rvs}}
\end{figure*}

As mentioned above, we are particularly interested in using our measured spectra as a means to explore the radial velocity (RV) precision of the comb-calibrated LHR approach. We determine frequency shifts (or radial velocities) from our measured spectra using a cross-correlation approach. In this approach, each telluric-corrected spectrum is compared to a template spectrum, and the observed frequency shift is determined as the shift that maximizes the cross correlation between the measured spectrum and the template. We construct a template for this procedure from the average of all spectra measured from 2023 September 21 to 2023 November 5. The template spectrum is smoothed using the procedure described by \cite{fredrick_thermal-light_2022} to remove high-frequency noise. 

After determining a frequency shift for each measured spectrum, we filter the measured RVs to eliminate measurements that may be biased by external effects (e.g. residual telluric contamination or clouds). Specifically, we restrict our measurements to times where the atmospheric airmass is less than 2.5 to mitigate residual telluric effects and we also eliminate spectra with large amplitude fluctuations that are indicative of variable atmospheric transmission (e.g. clouds). After applying this preliminary filtering, a noticeable diurnal pattern remains in our measured radial velocities. As we discuss below, we attribute this pattern to a pointing error induced by an asymmetry in the antenna pattern with respect to the center of the solar disk. We determine an empirical correction for the repeatable diurnal pattern based on a fit to the pattern averaged over the six-week dataset (see Section \ref{sec:pointingunc}), and we use this correction to detrend the measured radial velocities for each individual day. In a final step, we also test for correlations between our measured radial velocities and the temperature of the LHR instrumentation, which is located in an air-conditioned rooftop laboratory near the telescope. We eliminate days with a measurable temperature correlation that could indicate a potential bias in the measured RVs. Further details describing the pointing-induced diurnal pattern and temperature-induced uncertainty are discussed in Sections \ref{sec:pointingunc} and \ref{sec:tempunc}, respectively. Taken as a whole, our data filtering and pre-processing procedure eliminates 23 of the 46 days of solar observations, with 13 eliminated due to weather and an additional 10 removed due to temperature correlations.

Figure \ref{fig:rvs} shows our radial velocity measurements from 2023 September 21 to 2023 November 5. The figure shows the time series of each individual shift measurement (corresponding to individual 5~s LHR spectra) as well as the average shift for each day. The average shift is calculated as the uncertainty-weighted mean of each 5~s measurement over a given day. Section \ref{sec:uncert} describes our uncertainty estimates for both the individual measurements and the averaged shifts. 

We use the long-duration measurements to explore the RV precision of comb-calibrated LHR for timescales spanning hours to weeks. To better illustrate the performance over a single day, Figure \ref{fig:rvs}(c) and (e) shows the time series of radial velocities for data recorded on 2023 September 28 and 2023 October 17, which are representative of the types of systematic variations we find in the broader dataset. Panels (d) and (f) show the Allan deviation \citep{allan1966statistics} of the measured shifts, which provides information on the frequency precision as a function of averaging time, $\tau$. For both days, the Allan deviation indicates a frequency precision of $\sim$12~MHz for each 5~s measurement, equivalent to a radial velocity precision of $\sim$18~m/s. The Allan deviation shows two curves for both days, one for the ``raw" shift measurements without detrending, and the other after detrending to account for a repeatable diurnal pattern mentioned above. For both days, the precision in the raw shift measurements improves nearly with $\sqrt{\tau}$ to $\sim$1~MHz ($\sim$1.5~m/s) before the averaging is limited by pointing-induced frequency drift. 

Detrending the data to compensate for pointing-induced drift significantly improves the radial velocity precision over longer timescales. For the data on October 17, the detrended data averages nearly with $\sqrt{\tau}$ over the full measurement period, reaching a frequency precision of $\sim$300~kHz (or radial velocity precision of $\sim$45~cm/s). This precision is equivalent to a fractional frequency precision of a few parts in $10^{-9}$, splitting the $\sim$5~GHz line width by a factor of more than $10^4$. Detrending has less of an effect on the shifts measured on September 28. In this case, detrending improves averaging relative to the raw data, however the averaging still indicates instability on the timescale of $\sim$1000-2000~s. It is possible that this instability stems from weak temperature-induced fluctuations that occur over these timescales (see Section \ref{sec:tempunc}). Nonetheless, the radial velocity precision reaches $\sim$1~m/s despite the imperfect averaging. 

We also use the long-duration dataset to investigate the precision of our measurements over the full, six-week dataset. This long-term stability is most easily assessed through the averaged shift measurements shown in Figure \ref{fig:rvs}(b). In this case, the average shift measurements indicate a drift in the measured transition frequency by $\sim$310~kHz per day (the dashed line in \ref{fig:rvs}(b)). Although the center frequency of the \ion{Fe}{1} line is not expected to be static due to solar activity effects, RV drift at the level shown in Figure \ref{fig:rvs}(b) is likely due to an instrumental effect. It is difficult to explain a frequency drift at this level as being caused by our calibration or post-processing procedure (see discussion in Section \ref{sec:uncert} below). Instead, we find it more likely that this drift is also related to asymmetry in the antenna pattern and instability in the fiber coupling optics, which we discuss in more detail below.

\section{\label{sec:uncert}Description and Estimation of Uncertainty Sources}

A central goal of this effort is to use our long-duration solar measurements to investigate the RV precision of our LHR approach as well as to explore the principal uncertainty sources that impact our measurements. While our description of uncertainty sources is not intended as a final uncertainty budget, the estimates below capture our present understanding of the capabilities and performance of the comb-calibrated LHR. As such, this work will inform instrumentation modifications aimed at improving the approach in future applications. 

\subsection{\label{sec:photon}Photon Noise, Amplitude Fluctuations, and Other Random Uncertainties}

Random fluctuations in the measured radial velocities are primarily driven by amplitude noise in the measured spectra. The spectral signal-to-noise ratio is limited by the shot noise (photon noise) of the local oscillator (see Equation \ref{eq:snr}), and this noise manifests as random fluctuations in the measured radial velocities after the cross correlation process described above. While the photon noise is stationary, the resulting radial velocity noise is not as the signal level changes throughout the day due to airmass dependent losses. We limit our radial velocity measurements to airmass less than 2.5, and the increase in noise at that airmass relative to the value at zenith is minor ($\sim$8\%). 

Although photon noise is the dominant source of random uncertainty in our measurements, another important source stems from signal fluctuations within individual laser scans. Scan-to-scan changes of the signal amplitude effectively add a variable ``baseline" to individual spectra that can manifest as an apparent frequency shift relative to the template spectrum. We have effectively eliminated amplitude fluctuations induced by changes in the LO laser power by actively stabilizing the laser power, which remains stable at the level of one part in $10^4$ or better over a single 5~s laser scan. However, amplitude fluctuations can still arise due to changes in the atmospheric transmission over each 5~s scan. Our radial velocity measurements are particularly sensitive to these small amplitude fluctuations due to the limited bandwidth of our measurements, and we estimate that an amplitude fluctuation at the 1\% level is sufficient to induce an apparent shift of $\sim$10~MHz ($\sim$15~m/s). Our data processing procedure eliminates measurements made under cloudy conditions (where amplitude variations are most significant), however it is possible that spectra measured under imperfect atmospheric conditions could exhibit amplitude-induced radial velocity fluctuations. 

To quantify the random uncertainty in our RV measurements, we calculate both ``local" and ``global" estimates of the standard deviation of our measured radial velocities. The global estimate calculates the standard deviation of the time series of radial velocities measured each day, which are detrended using a low frequency smoothing function to avoid including drift and systematic uncertainties that are estimated individually. This global estimate does not account for non-stationary processes, and may underestimate the random uncertainty in time periods where (for example) variable atmospheric transmission increases fluctuations in the measured radial velocities. To capture these non-stationary effects, we also calculate a ``local" estimate of the standard deviation of the measured radial velocities in three minute time bins. This local estimate captures increased radial velocity scatter over short time periods. We set the final estimate of the random uncertainty for each individual radial velocity measurement as the greater of the local and global estimates in the corresponding time bin.

Lastly, we note that other experimental effects can contribute to random fluctuations in our radial velocities besides photon noise and signal variations. These effects include fluctuations in the repetition rate of our frequency comb and components of our frequency calibration process, which we describe below.

\subsection{\label{sec:combunc}Frequency Reference}

An advantage of our approach is the traceability of the frequency axis to absolute standards via the optical frequency comb.  Such frequency combs have been rigorously evaluated to be capable of intrinsic uncertainty at (and below) 1 part in $10^{19}$\citep{ma2004optical}. However, uncertainties that might arise from the particular manner in which we use the comb merit a careful analysis. From Equation \ref{eq:comb}, the frequency of each comb calibration ``tick" is specified by the integer index of the tick ($n$), the repetition rate of the frequency comb ($f_r$) and the carrier-envelope offset frequency ($f_o$). As described above, we employ a reference CW laser locked to our frequency comb to identify and track the mode number of a specific calibration tick in all of our measurements. To determine the mode number, we measure the frequency of the CW laser and solve Equation \ref{eq:comb} for $n$ using the known values of the repetition rate, the carrier-envelope offset frequency, and the CW laser lock offset. We measure the frequency of the CW laser using a Bristol 621 wavelength meter with an accuracy of 0.2 pm at 1565~nm ($\sim$36 MHz). We found the measured CW laser frequency to be within $\sim$7~MHz of its expected frequency based on the known frequency comb parameters and lock offset. As such, we assume no uncertainty associated with the determination of $n$.

The remaining uncertainty associated with the frequency comb reference stems from uncertainty in the repetition rate and carrier-envelope offset frequency. Both parameters are referenced to a NIST-calibrated hydrogen maser with uncertainty $<$1~mHz. The uncertainty in the carrier-envelope offset frequency is additive while the uncertainty in the repetition rate is multiplicative (see Equation \ref{eq:comb}). As such, the uncertainty in the $n^{\rm th}$ calibration tick is well approximated as $n \delta f_r$, where $\delta f_r$ is the uncertainty in the repetition rate. We measure $\delta f_r$ to be $\sim$500~\textrm{$\mu$}Hz over timescales of more than \textrm{$10^4$}~s, which is equivalent to a calibration tick uncertainty of $\sim$750~Hz in our measurement window at 1565.3~nm ($n=1,532,232$).

\subsection{\label{sec:calunc}Frequency Axis Uncertainty}

While the uncertainty of the underlying frequency comb is effectively negligible, our use of the comb requires the transformation of the temporal scan of the LO laser onto the comb-referenced frequency grid. To accomplish this, we fit each calibration tick with a Gaussian function to determine its centroid, and then interpolate those points to apply the calibration to the full laser scan. Both the centroiding and interpolation components of this process contribute uncertainty. We estimate the uncertainty of the centroiding procedure as the standard error of the best-fit centroid following a Gaussian fit to our measured calibration ticks in the frequency domain. We find the average uncertainty in the tick centroid to be $\sim$38~kHz. 

We use the Lagrange remainder formula to estimate the uncertainty associated with the interpolating function that fits the measured calibration points \citep{epperson2013introduction}. For the 2\textsuperscript{nd} order polynomial interpolation employed here, this uncertainty is bounded as 
\begin{equation}\label{eq:interpmax}
|\epsilon|_{max}\leq\frac{h^3}{9\sqrt{3}}\times \textrm{max}|f^{(3)}(t)|
\end{equation}
where $h$ is the temporal spacing between calibration ticks and $f^{(3)}(t)$ is the 3\textsuperscript{rd} derivative of the time-to-frequency calibration function (which we estimate numerically). Using this approach, we estimate the uncertainty associated with the interpolation process to be $\sim$1~kHz (neglecting the edges of our measured spectra where the laser scan is nonlinear and the derivatives are large). 

The final uncertainty component associated with our calibration process is the uncertainty in our method for correcting for the group delay between the solar and comb-calibration channels. As described above, our group delay correction relies on a measurement of the phase response of the final low pass filter and amplification stage in both the solar and comb calibration branches. We multiply the measured solar and comb calibration signals by the inverse of the corresponding phase response to remove the differential group delay between the two signals. This process is critical for accurate frequency measurements, as differential delay between the solar and calibration signals leads to a frequency shift after the time-to-frequency calibration process. 

Our group delay correction process uses the average of 100 individual measurements of the phase response in both branches to correct for the group delay. To estimate the uncertainty in this process, we apply the group delay correction to a simulated absorption transition for each of the 100 filter response measurements, and determine the resulting frequency shift after the calibration process relative to a simulated spectrum with no group delay. The result is 100 measurements of group delay-induced frequency shift in our measured spectra, and we estimate the uncertainty as the standard error of the mean shift. This estimate yields an uncertainty of $\sim$8~kHz associated with group delay correction.

\subsection{\label{sec:pointingunc}Solar Tracking and Antenna Pattern}

The Sun's rotational velocity presents a significant opportunity for bias in our RV measurements since any asymmetric coupling of light from across the solar disk introduces a rotational Doppler shift that affects the apparent center of the measured transition. In practice, these inadvertent rotational Doppler shifts can be induced by imperfect solar tracking or any asymmetry in the telescope's antenna pattern across the solar disk. 

\begin{figure}[t!]
\includegraphics[width=\linewidth]{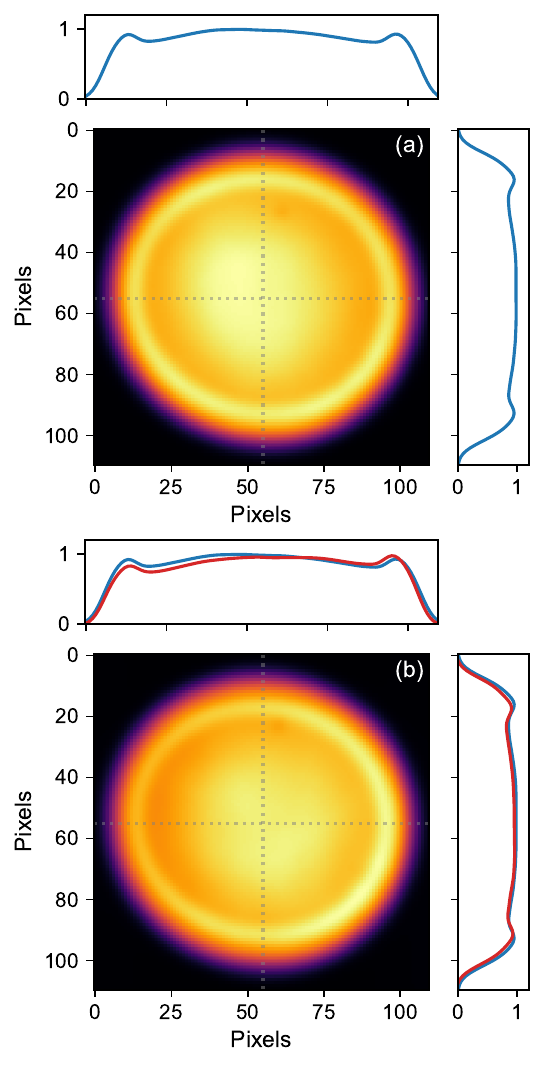}
\caption{The LHR antenna pattern imaged with a beam profiler. Panel (a) shows the antenna pattern measured before the start of the six week dataset along with horizontal and vertical cross sections measured along the center line. Panel (b) shows the pattern measured after six weeks of continuous observations. Comparison of the cross sections measured before data collection (blue) and after (red) shows a clear asymmetry that has developed due to relaxation of the fiber coupling optics.
\label{fig:antenna_pattern}}
\end{figure}

\begin{figure}[t]
\includegraphics[width=\linewidth]{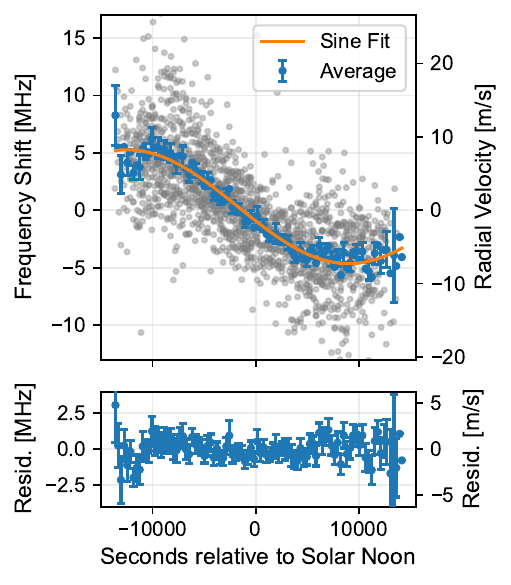}
\caption{Diurnal pattern in the measured radial velocities. Gray points show measured radial velocities for each day plotted on the same axis as a function of time relative to solar noon. The mean RV is subtracted from each day and the RVs are averaged in 300~s bins to reduce noise. Blue points show the average RV in each 300~s bin across the full, six-week dataset along with the best-fit sine wave. Error bars indicate the standard error of the mean RV in each bin. The bottom panel shows the residuals (measurement - model) between the average RV and the fit result. 
\label{fig:diurnal_pattern}}
\end{figure}

Figure \ref{fig:antenna_pattern} shows an image of the LHR antenna pattern measured before and after the six-week observation period. Although subtle, Figure \ref{fig:antenna_pattern}(b) shows a clear asymmetry that developed over the six-week dataset. As discussed above, the measured RVs exhibit a repeatable diurnal pattern that we attribute to a rotational Doppler shift introduced by this asymmetry. Figure \ref{fig:diurnal_pattern} shows this diurnal pattern in the measured radial velocities. The gray points show all measured RVs from 2023 September 21 to 2023 November 5 plotted on the same axis as a function of time relative to solar noon. The daily mean has been subtracted, and the measurements are binned by five minutes to reduce noise. Blue points show the average RV in each five-minute bin (again with the daily mean subtracted), which clearly shows the repeatable, $\sim$10~MHz peak-to-peak drift in the daily RV measurements. The sinusoidal shape of the drift pattern is consistent with what we expect for shifts induced by asymmetric coupling of light from across the solar disk.

To account for this observed drift, we fit the mean diurnal pattern with a sine wave and use the best-fit curve to detrend the measured shifts for each individual day. This fit result is also shown in Figure \ref{fig:diurnal_pattern}. The best-fit curve reproduces the observed drift to within a few MHz. We estimate the uncertainty of this correction using the standard error of the mean RV in each time bin (the error bars on Figure \ref{fig:diurnal_pattern}). This uncertainty increases for measurements made farther from solar noon, as we have fewer measurements in those time periods to constrain the shape of the diurnal pattern and the associated fit result. Notably, this estimate only covers the uncertainty associated with our empirical correction and does not consider whether the model provides a complete correction for pointing errors. As such, we estimate this uncertainty using $3\sigma$ coverage to provide a more conservative estimate of the pointing-induced uncertainty. We discuss prospects for improving the stability of our antenna pattern and solar tracking optics in Section \ref{sec:discussion}.

\subsection{\label{sec:tempunc}Temperature-Induced Frequency Shifts}

All of the radiometric components (i.e. components involved in the conversion of optical power to voltage) are housed in insulated enclosures and mounted to water-cooled breadboards that are temperature stabilized using a recirculating chiller ($\pm$0.05 \degree C stability). We monitor the temperature of critical components using thermistors mounted to each component.

Despite these temperature stabilization measures, we have observed residual correlations between our measured radial velocities and the temperatures measured on various experimental components. These correlations are likely exacerbated by the relatively large temperature fluctuations in the rooftop laboratory used for these measurements, which reach approximately $\pm$0.75 \degree C for the ambient lab temperature and $\pm$0.1 \degree C within our insulated enclosures despite the active temperature stabilization described above. 

\begin{figure*}[t]
\includegraphics[width=\linewidth]{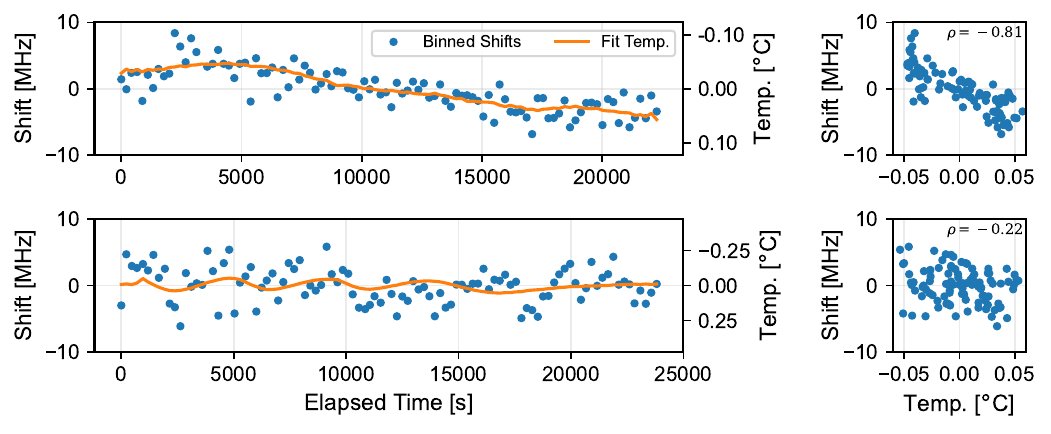}
\caption{Temperature-induced frequency shifts in the measured data. The top panel compares the measured RVs for 2023 October 20 to the temperature measured on the mixer in RF power detection circuit. The RVs have been averaged in $\sim$250~s bins to reduce noise, and the measured temperatures have been scaled and shifted to best fit the measured frequency shifts. There is a clear correlation between the shifts and temperatures ($\rho=-0.81$). The bottom panel shows the same comparison for the RVs measured on 2023 October 13, where no clear correlation is observed.  
\label{fig:tempcorr}}
\end{figure*}

Our data processing procedure identifies and omits days that exhibit a significant correlation between measured radial velocities and laboratory temperatures. To identify these correlations, our processing code attempts to scale and shift the measured temperatures to match the observed frequency shifts. This fitting process includes two free parameters: a scaling factor (MHz/\degree C) that scales the small temperature fluctuations to our measured shifts, and a temporal offset that aligns the measured temperatures and frequency shifts in time. Figure \ref{fig:tempcorr}(a) and (c) show the result of this fitting process for two representative days. Notably, the temporal offset in the fitting process makes this approach agnostic to the specific component or components driving the temperature sensitivity because the temperature of all components differ only by a scaling factor and temporal delay. As such, we use the temperature of the RF mixer to assess correlations and evaluate the associated uncertainty. 

The fit result in Figure \ref{fig:tempcorr}(a) shows a clear correlation between the measured frequency shifts and the scaled and shifted mixer temperature for the data collected on 2023 October 20, while panel (c) shows no clear correlation in data from October 13. Panels (b) and (d) plot the frequency shifts as a function of the mixer temperature, which provides another view to assess the correlation. We quantify the degree of correlation by calculating the Pearson correlation coefficient ($\rho$) between the measured shifts and the scaled, shifted temperatures. A value $\rho=1$ indicates a perfect linear correlation, and $\rho=0$ indicates no correlation. We assume days with $\rho>0.5$ to be indicative of a significant temperature correlation that may bias our measured radial velocities, and we omit those days from our results.

The process described above also provides a means to estimate the temperature-induced frequency uncertainty of our measurements. We estimate this uncertainty based on the average temperature scaling factor measured for the days with a significant temperature correlation ($\rho>0.5$). We estimate the temperature-induced frequency uncertainty to be this average scaling factor ($\sim$77 MHz/\degree C) multiplied by the standard deviation of the measured mixer temperature for each day. \newline

\subsection{\label{sec:telluricunc}Telluric Correction}

The measured solar transition overlaps with several atmospheric absorption transitions from both $\mathrm{H_2^{16}O}$ and $\mathrm{HD^{16}O}$. This telluric absorption affects the perceived center of the measured solar transition, and the resulting frequency shift varies as the telluric absorption depth changes with airmass. To account for this effect, our data processing procedure fits and subtracts the telluric absorption using a HITRAN2020-based absorption model \citep{gordon_hitran2020_2021} and a multi-layer atmospheric model. 

Telluric absorption in the 1565~nm region is weak and well below the noise level of our individual measurements at low airmass. As such, our approach uses spectra measured each morning (when airmass is high and telluric absorption is strong) to construct a template spectrum that is used for telluric correction. We fit the template spectrum with our multi-layer model to constrain the telluric model, and use the best-fit model to subtract a telluric spectrum (scaled by airmass) from each individual measurement. Airmass is calculated using a Python wrapper for the NREL Solar Position Algorithm \citep{reda_solar_2004, f_holmgren_pvlib_2018}. 

\begin{figure}[t]
\includegraphics[width=\linewidth]{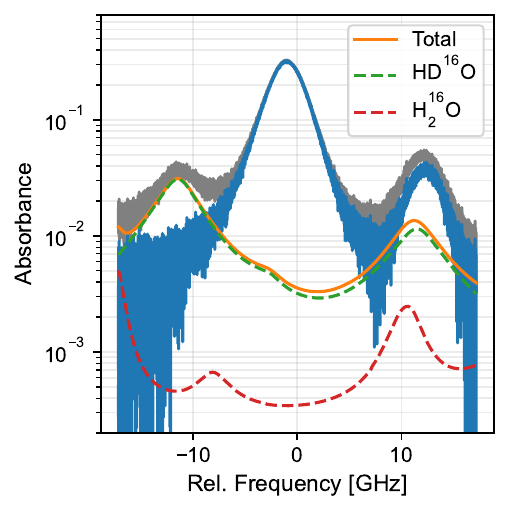}
\caption{Solar spectrum measured on the morning of 2023 October 4 before and after telluric correction. The spectrum before telluric correction is shown in gray, and the telluric-corrected spectrum is shown in blue. The spectrum is an average of 250 individual measurements ($\sim\textrm{20 minutes}$) with an average airmass of 2.5.
\label{fig:tellurics}}
\end{figure}

Owing to the weak telluric absorption, we don't fit $\mathrm{H_2^{16}O}$ and $\mathrm{HD^{16}O}$ individually and instead fix their ratio according to their known natural abundance \citep{hill2016hitranonline}. Additionally, due to the airmass scaling, our approach does not account for temporal or spatial variations in the water column throughout each day, however we have not observed indications of these effects in our measurements. Figure \ref{fig:tellurics} shows an example fit result as well as a measured solar spectrum before and after telluric correction. 

The telluric correction procedure can contribute uncertainty in our measured radial velocities due to residual telluric absorption remaining after subtracting the best-fit model, or due to changes in the telluric absorption throughout each day that are not captured by our telluric correction approach. To estimate this uncertainty, we simulated the addition of varying levels of residual telluric absorption to a telluric-free, high-SNR reference spectrum. We then measured the resulting bias in the RV measurements. For typical levels of residual telluric contamination observed in our measured spectra ($\lesssim 1$\%), the magnitude of the resulting bias is approximately 20~kHz.

\subsection{\label{sec:combining}Combining Uncertainty Contributions}

We add the uncertainty estimates described above in quadrature to estimate an uncertainty for each individual radial velocity measurement. For a single measurement (5~s), this uncertainty is dominated by random fluctuations at the level of $\sim$12~MHz driven by shot noise and random amplitude fluctuations (see Section \ref{sec:photon}). The average radial velocity shift for each day is determined as an uncertainty-weighted average of the individual measurements. The uncertainty in the mean is again the quadrature sum of our uncertainty estimates with the random uncertainty component given by the weighted standard error of the mean. In this case, the uncertainty in the averaged radial velocities is limited by the uncertainties associated with temperature-induced frequency shifts ($\sim$1.7~MHz) and our correction for the diurnal drift driven by asymmetry in the LHR antenna pattern ($\sim$2~MHz). Lastly, while this section discusses the primary uncertainty components that impact our measurements, we do not consider other potential contributions such as differential extinction across the solar disk, the effect of atmospheric emission lines, or polarization effects, which could be the subject of future studies.

\section{\label{sec:discussion}Discussion, Prospects and Future Modifications}

Our investigation of uncertainty sources makes it clear that temperature sensitivity and instability in the LHR antenna pattern represent the principal limitations of our comb-calibrated LHR system in its current form. These two effects contribute systematic uncertainties of $\sim$1.7 and $\sim$2~MHz (respectively), approximately two orders of magnitude higher than the uncertainties contributed by the frequency comb reference and calibration procedure (10s of kHz or less). As such, realizing the full potential of the frequency comb calibration for precision solar spectroscopy will require future modifications to improve the stability of the antenna pattern and reduce sensitivity to temperature fluctuations. 

Asymmetry in the LHR antenna pattern manifests as both a repeatable diurnal pattern in the measured radial velocities as well as a long-term drift over the six week dataset. This asymmetry is caused by transverse misalignment of the fiber optic collimator relative to the beam shaping optic that forms the flat-top profile. If the antenna pattern is asymmetric, more light is coupled from a specific portion of the solar disk, leading to a rotational Doppler shift that varies throughout the day as the Sun's rotation axis changes relative to the antenna pattern (e.g. Fig. \ref{fig:diurnal_pattern}). We also attribute the long-term drift in our measured RVs (Fig. \ref{fig:rvs}(b)) to asymmetry in the antenna pattern. Day-to-day changes in the optical alignment can vary the degree of asymmetry in the antenna pattern and thus also the magnitude of the rotational Doppler shifts that bias the measured RVs. Long-term drift could also arise from day-to-day changes in the orientation of the solar rotation axis relative to the antenna pattern. However, since we observe changes in the antenna pattern after the six-week dataset (Fig \ref{fig:antenna_pattern}), we assume that changes in the optical alignment are the primary driver of long-term RV drift in our measurements. 

While we can partially compensate for the diurnal drift induced by the antenna pattern (see Sec. \ref{sec:pointingunc}), it is difficult to rigorously account for the long-term drift in the measured RVs since the antenna pattern cannot be measured \textit{in situ}. As such, future modifications will pursue more robust fiber-coupling and beam shaping optics that are less sensitive to environmental perturbations (e.g. thermal cycling and vibrations) that affect the alignment of our solar tracking optics. One such modification could include the substitution of an integrated optic that enables fiber coupling and beam shaping in a single component, eliminating the potential for misalignment. An alternative approach could modify the solar telescope to create a larger antenna pattern. A larger antenna pattern would reduce the SNR, but could also significantly reduce sensitivity to the exact shape of the antenna pattern and sensitivity to misalignment.

In addition to instability in the antenna pattern, we also show that temperature-induced RV fluctuations impact the long-term stability of our measurements. Our preliminary analysis of this effect indicates that the correlations between the measured temperatures and RVs are intermittent but can impact the measured frequency shift by up to 77~MHz/$\degree$C. However, despite recording the temperature across eight separate components (including fiber optics, RF electronics and photodetectors), we are not able to identify a single physical mechanism that explains correlations between RVs and temperature. It is possible that changes in temperature affect the gain in the RF power detection chain or induce chromatic transmission variations in our fiber optics. Both effects could induce a variable baseline in our measured spectra that manifests as a frequency shift. Ongoing work is focused on mitigating these temperature correlations by improving temperature stability in our instrument and by constraining the mechanism by which temperature fluctuations affect the measured RVs.  

More broadly, in addition to informing future modifications to our instrument, our results also demonstrate the benefits of comb-calibrated LHR for precision solar spectroscopy. This study shows that the LHR approach is capable of daily Sun-as-a-star observations over a six week period, and that the radial velocity precision can routinely reach 1~m/s or better within a single day. Further, although not discussed in detail in this paper, we show that the LHR approach also enables line shape measurements with high signal to noise ratio ($\sim$2,600), high spectral resolution ($\sim800,000$), and absolute frequency accuracy. These results demonstrate the potential of comb calibrated LHR as a tool for precision solar spectroscopy and further motivate future efforts to improve the technique for long duration studies of solar variability.

\section*{Funding}
This work was supported by the NIST Innovations in Measurement Science program, NIST
financial assistance award 70NANB18H006, the NASA Astrophysics Division (grant \#80NSSC21K2006), and the W. M. Keck Foundation. R.C. acknowledges support from the National Academies NRC
Research Associateship Program.

\section*{Acknowledgements}
We thank Drs. Nate Newbury and Ian Coddington for sharing rooftop lab space to accommodate this project and Drs. Nazanin Hoghooghi, Takuma Nakamura and Daniel Slichter for providing useful feedback. This work is a contribution of NIST and is not subject to copyright in the U.S. 

Certain equipment, instruments, software, or materials are identified in this paper in order to adequately specify the experimental procedure. Such identification is not intended to imply recommendation or endorsement of any product or service by NIST, nor is it intended to imply that the materials or equipment identified are necessarily the best available for the purpose.

\bibliography{manuscript}{}
\bibliographystyle{aasjournal}

\end{document}